\documentclass[twocolumn,abs,prb,showpacs]{revtex4-1}

\usepackage{graphicx}
\usepackage{dcolumn}
\usepackage{float}
\usepackage{amsmath}
\usepackage{bm}

\newcommand{\comment}[1]{}

\begin{document}


\title{Optical conductivity of twisted bilayer graphene}

\author{C.J. Tabert}
\author{E.J. Nicol}
\affiliation{Department of Physics, University of Guelph,
Guelph, Ontario N1G 2W1 Canada} 
\affiliation{Guelph-Waterloo Physics Institute, University of Guelph, Guelph, Ontario N1G 2W1 Canada}
\date{\today}

\begin{abstract}
{We calculate the finite-frequency conductivity of bilayer graphene with a relative twist between
the layers.
The low frequency response at zero doping shows a flat conductivity with value twice that of the monolayer case
and at higher frequency a strong absorption peak occurs.
For finite doping, the low frequency flat absorption is modified into a 
peak about zero frequency (the Drude response) accompanied by an interband edge which results
from the transfer of spectral weight from interband to intraband absorption due to Pauli blocking. 
If the system is doped sufficiently such that the chemical potential reaches beyond the low-energy saddle point in the twisted bilayer
band structure, a strong low frequency absorption peak appears at an energy
related to an effective interlayer hopping energy, which may be used to
identify this parameter and confirm the existence of the saddle point which gives rise to a low energy van Hove singularity
in the electronic density of states. 
}
\end{abstract}

\pacs{78.67.Wj,78.30.-j,78.20.Ci,81.05.ue}


\maketitle


Graphene remains a material of considerable promise both for technological applications and 
for revealing unusual and unexpected physics.\cite{Neto:2009,Abergel:2010}
 Key to this enterprise is the ability to 
manipulate its band structure and change the Fermi level through charge doping by electrons or holes. In the former, the layering of 
graphene sheets in various stacking arrangements can produce very different
dispersions at low energy, such as quadratic in the Bernal-stacked bilayer
versus linear in the monolayer. Recently, it has been noted that 
layers of graphene with a rotational misorientation can give rise 
to surprising behavior
at low energy. Indeed for small rotational angles of a bilayer of graphene,
it is predicted that the low energy dispersion will be linear not unlike the monolayer\cite{Santos:2007}
and a low energy van Hove feature will also appear in the density of states.\cite{Li:2009,Luican:2011,Brihuega:2012}
Moreover, the Fermi velocity $v_F$ is reduced for small rotation angles\cite{Santos:2007} until
localization appears to set in.\cite{Trambly:2010,San-Jose:2012} These features have been the subject of
a number of theoretical works, such as Refs.~\cite{Santos:2007,Bistritzer:2011,Trambly:2010}, and have been verified by experimental groups
performing scanning tunneling microscopy (STM) measurements,\cite{Li:2009,Luican:2011,Brihuega:2012}
although controversy remains as will be discussed below.
 \begin{figure}[h!]
\includegraphics[width=1.0\linewidth]{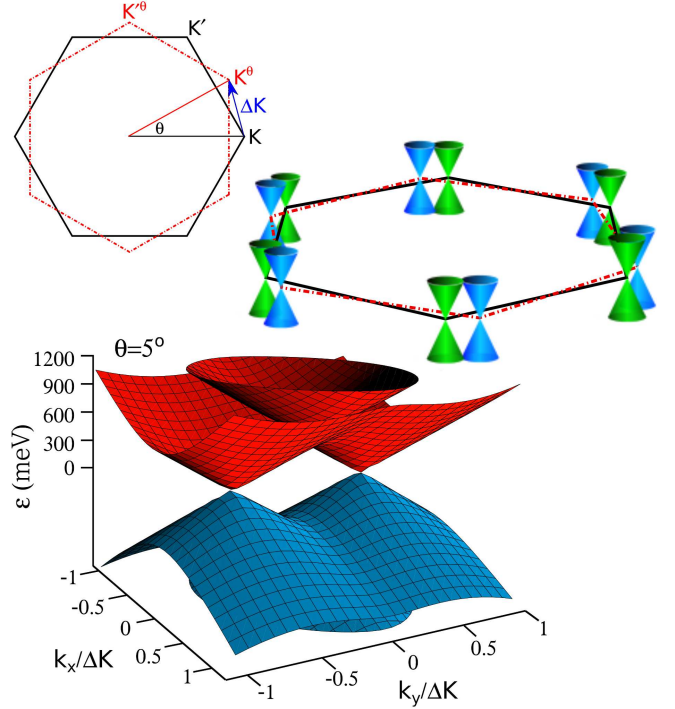}
\caption{(Color online) Top: The First Brillouin Zones of two graphene
layers where one layer is twisted (dashed red) relative to the other  
(solid black).  The two $K$ points, 
$\bm{K}$ and $\bm{K}^\theta$, are separated by $\Delta \bm{K}$. 
Middle: The relative rotation shifts the Dirac cones in one layer 
relative to those in the other layer, as shown schematically with
blue and green cones for each layer, respectively. 
Bottom: The low-energy band structure about the midpoint between the two Dirac cones shifted relative to the $K$ point due to a twist angle of $\theta=5^\circ$. 
}\label{fig1}
\end{figure}

The origin of these features is found in the details of the modified energy dispersion.
In twisted bilayer graphene, 
the Dirac cones at the $K$ points of the Brillouin zone 
in one layer undergo a relative rotation to
those in the other layer (as shown in Fig.~\ref{fig1}). The 
two Dirac cones which are slightly shifted relative to each other then 
overlap and the band structure is reconstructed to form a saddle point 
in between. The final modified band structure around the midpoint
of the split Dirac point near $K$ is illustrated in the lower
part of Fig.~\ref{fig1} for the two lowest energy bands in the model discussed here.
At low energy, the slope of the linear band structure  gives a Fermi velocity 
which is reduced from the graphene monolayer value, and the saddle point in between produces a van Hove 
singularity (VHS) in the density of states at low energy. One might have naively considered  the
graphene sheets to have decoupled with the relative rotation, but the presence of the 
renormalized Fermi velocity and VHS point to a different interpretation.
In spite of a growing literature on twisted bilayer graphene, there have
been few experiments which have verified these unusual results. In particular,
the VHS saddle point and renormalized velocity are not seen in some angle-resolved photoemission
experiments (ARPES)\cite{Hicks:2011}, but 
these features are very clear in the STM\cite{Li:2009,Luican:2011,Brihuega:2012} and some ARPES measurements do confirm the VHS\cite{Ohta:2012}.
The predicted behavior of the Landau levels in twisted bilayer graphene\cite{deGail:2011,Choi:2011} 
as measured through the quantum Hall effect has not been seen in rotated layers,\cite{Lee:2011} but Raman
spectroscopy of the rotational-angle dependent graphene 2D peak\cite{Kim:2012} points to the existence
of the VHS in the band structure. Further experiments are required to resolve this situation and
indeed other spectroscopies should be brought to bear on this question.
The interplay of theory and experiment for the dynamical conductivity of graphene
and bilayer graphene has been quite successful in the past\cite{Ando:2002,Gusynin:2006a,Gusynin:2006,Peres:2006,Falkovsky:2007,Stauber:2008,Abergel:2007,Nilsson:2008,Nicol:2008,Koshino:2009,Zhang:2008,Li:2008,Kuzmenko:2008,Wang:2008,Nair:2008,Mak:2008,ZLi:2009,Kuzmenko:2009a,Kuzmenko:2009b,Orlita:2010} and consequently, 
we provide here the theoretical calculation for  
the dynamical conductivity of twisted bilayer graphene, illustrating
how the VHS will be manifest in this experiment.


A literature has developed for modelling misorientated bilayer graphene. A 
popular model for describing the state of twisted bilayer graphene is the
continuum model put forth by Lopes dos Santos and coworkers\cite{Santos:2007}
 where the Hamiltonian is written as\cite{deGail:2011}
 \begin{equation}\label{HamiltonianTW}
H(\bm{k})=\left(\begin{array}{cc} 
H_0(\bm{k}+\Delta\bm{K}/2)& H_\perp \\ 
H^\dagger_\perp& H_0(\bm{k}-\Delta\bm{K}/2)
\end{array}\right),
\end{equation}
with
 \begin{equation}\label{HamiltonianG}
H_0(\bm{k})=
\left(\begin{array}{cc} 
0 & f^*(\bm{k}) \\ 
f(\bm{k})&0\end{array}\right),
\end{equation}
where $f(\bm{k})=\hbar v_F(k_x+ik_y)$
and
 \begin{equation}\label{Hamiltonianperp}
H^0_\perp(\bm{k})=\tilde t_\perp
\left(\begin{array}{cc} 
1 & 1 \\ 
1&1\end{array}\right),
H^\pm_\perp(\bm{k})=\tilde t_\perp
\left(\begin{array}{cc} 
e^{\mp i\phi} & 1 \\ 
e^{\pm i\phi}&e^{\mp i\phi}\end{array}\right),
\end{equation}
where $\phi=2\pi/3$ and $\tilde t_\perp$ is an angle-dependent 
interlayer
hopping parameter typically quoted as being between 100-150 meV.
Here, the rotation between the layers is captured by the monolayer graphene 
Hamiltonian of Eq.~\eqref{HamiltonianG}, where the argument is replaced by 
$\bm k\pm\Delta \bm K/2$ with $\Delta \bm K=\bm K-\bm K_\theta$. The expansion in $k$
is taken about the midpoint between the two shifted Dirac cones as shown
in Fig.~\ref{fig1}.
The interlayer hopping terms represent the dominant Fourier amplitudes in
the interlayer hopping as described in Ref.~\cite{Santos:2007}. This Hamiltonian and also ab initio 
and tight-binding methods have been 
employed
by various authors to determine the band structure and density of electronic
states. However, when including
a magnetic field, some authors\cite{deGail:2011,Choi:2011,Lee:2011} 
have examined the Landau level structure using a low-energy effective 
Hamiltonian where the interlayer hopping is taken as the standard Bernal bilayer form:\cite{deGail:2011,Choi:2011} 
 \begin{equation}\label{Hamiltonianperpeff}
H^{\rm eff}_\perp(\bm{k})=\tilde t_\perp
\left(\begin{array}{cc} 
0 & 0 \\ 
1&0\end{array}\right).
\end{equation}
In zero magnetic field, this leads to an analytic expression
for the low energy dispersion given by
\begin{align}\label{Energy}
\varepsilon^2_\alpha(\bm{k})&=\frac{1}{2}\left(\tilde t_\perp^2+\varepsilon_G^{+\,2}+\varepsilon_G^{-\,2}+(-)^\alpha\Gamma\right),\notag\\
\Gamma&=\sqrt{\left(\tilde t_\perp^2+\varepsilon_G^{+\,2}+\varepsilon_G^{-\,2}\right)^2-4\,\varepsilon_G^{+\,2}\,\varepsilon_G^{-\,2}},
\end{align}
 where $\alpha=1$ and 2 and $\varepsilon_G^\pm=|f(\bm{k}\pm\Delta\bm{K}/2)|$.  
In these works, extensive arguments for the validity of the approximation have
been given, including that the Hamiltonian remains in the same topological
universality class, preserves the chirality of the wavefunctions, and exhibits
a low energy band structure, similar to the other methods, including split
Dirac points and saddle points giving rise to the low energy
 VHS in the DOS. We have
examined this latter form and made comparisons to band structure of both
the G=0 approximation of the Lopes paper and with their full numerical results
presented in that paper and find reasonable agreement between the two approaches. As the low energy effective Hamiltonian is much more tractable
for a calculation of the optical properties using Green's functions and the
Kubo formula, we proceed with this effective Hamiltonian as used
recently by other authors\cite{deGail:2011,Choi:2011,Lee:2011} in the spirit of 
capturing the essence of the effect of rotational misalignment on the
finite frequency conductivity. Note that this approach will not be appropriate
for very small twist angles,
 where localization effects appear to set in, but should be suitable 
for\cite{Choi:2011}
$3^\circ\lesssim \theta\lesssim 10^\circ$ which is the region in which 
experiments have been performed. For much larger rotation angles it has been 
argued\cite{Lopes:2012,Mele:2012} that the model of Mele\cite{Mele:2010} should be used. This latter model has been used for the calculation of magneto-optics\cite{Apalkov:2011}.

\begin{figure}[h!]
\includegraphics[width=1.0\linewidth]{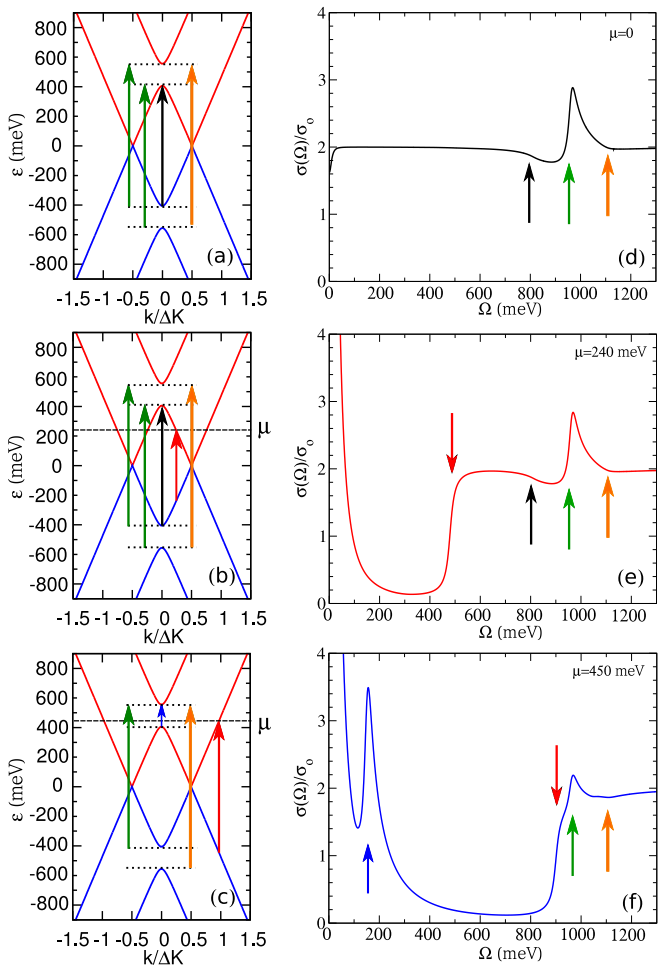}
\caption{(Color online)
(a)-(c) Band structure along a line connecting the two $K$ points $\bm{K}$ and $\bm{K}^\theta$
for $\theta=5^\circ$. In each successive frame, the chemical potential is changed from (a) $\mu=0$,
(b) 240 meV, (c) 450 meV, and various significant optical transitions are indicated which
give rise to structure in the conductivity curves shown to the right. (d)-(e) The finite
frequency longitudinal conductivity at zero temperature for varying chemical potential. The arrows
indicate the transitions shown in the band structure to the left. $\sigma_0=e^2/4\hbar$.}
\label{fig2}
\end{figure}

In Fig.~\ref{fig1}, we show the band structure evaluated from this approach for
an angle of 5$^\circ$. A cut of this band structure along the line connecting
the shifted Dirac cones is shown in Fig.~\ref{fig2}. Typically the interlayer
hopping $\tilde t_\perp(\theta)$ depends on angle and the specific value varies
in the literature. Consequently, to make 
our calculation
more applicable, we chose a value of $\tilde t_\perp(\theta)=150$ meV in our approach
to give a band structure with a $k=0$ saddle point energy and
upper band minimum to approximately match the energy scales found
in Ref.~\cite{Brihuega:2012} from ab initio and tight-binding calculations, and 
also confirmed therein by 
experimental data.  With this simplified model, we capture the essential
features found in the more numerical approaches: a linear dispersion at
low energy at each of the two shifted Dirac points and a saddle point in the band structure
at low energy. This should allow us to examine the signatures of these
features in the optical properties, at least at a qualitative level.

 For a calculation of the dynamical conductivity, we follow the method based on 
many-body Green's functions which is shown in the work by 
Nicol and Carbotte\cite{Nicol:2008} and Tabert and Nicol\cite{Tabert:2012} 
for the cases of AB- and AA-stacked bilayer graphene, respectively.  
Specifically,
 we can determine the Green's function $\hat{G}(z)$ from $\hat{G}^{-1}(z)=z\hat{I}-\hat{H}$ 
along with its spectral representation
$\hat{G}_{ij}(z)=\int_{-\infty}^\infty d\omega\,\hat{A}_{ij}(\omega)/[2\pi(z-\omega)]$.
Using the Kubo formula,\cite{Mahan:1990} where the conductivity is written in terms of the 
retarded current-current correlation function, we can express the real part of the finite 
frequency longitudinal conductivity, at zero temperature and for photon energy $\Omega$, as\cite{Nicol:2008}
\begin{equation}\label{CondIntegral}
\sigma(\Omega)=\frac{2\,e^2}{\Omega}
\int\,
\text{Tr}\left[\hat{v}_x\hat{A}(\bm{k},\omega+\Omega)\hat{v}_x\hat{A}(\bm k,\omega)\right],
\end{equation} 
where
\begin{equation}
\int \equiv \int_{|\mu|-\Omega}^{|\mu|}\frac{d\omega}{2\pi}\int\frac{d^2 k}{(2\pi)^2}\,
\end{equation}
with the $\bm k$ integration
over a region containing two shifted Dirac points. Here,
$\mu$ is the chemical potential and $\hbar\hat v_{x}=\partial \hat H/\partial k_x$. While the conductivity shown in Eqn.~\eqref{CondIntegral} is for $\sigma_{xx}$, it is the
   same for $\sigma_{yy}$, and is therefore isotropic. The velocity operator 
   brings in only the transport associated with a single graphene sheet, 
   while the reconstructed band structure enters only through the energies 
   in the spectral functions.


We now present the longitudinal conductivity which we obtain by a numerical evaluation of Eqn.~\eqref{CondIntegral} 
with the appropriate substitutions of the calculated spectral functions. The calculation follows steps similar to
those already in the literature\cite{Nicol:2008,Tabert:2012,Stille:2012} and the final expressions are lengthy and so we refrain from
repeating them here.  For the delta functions that appear in our expressions for the spectral functions, 
we use the Lorentzian representation $\delta(x)\to(\eta/\pi)/[\eta^2+x^2]$ with a broadening parameter $\eta=7$ meV
in order to do the numerical work.  
This broadening parameter manifests itself in the conductivity as an effective transport scattering rate of $1/\tau_{imp}=2\eta$ due to the 
convolution of two Lorentzians in the conductivity formula that result from the multiplication of the spectral functions.  

In Fig.~\ref{fig2}, we show the essence of the optical conductivity in twisted bilayer graphene within the model used here.
Plots of the conductivity at various dopings, marked by $\mu$, are exhibited  in Fig.~\ref{fig2}(d)-(f). 
For comparison the band structure is shown at the left in Fig.~\ref{fig2}(a)-(c) indicating the prominent absorptive
transitions
identified by arrows on the plots at the right. Note that the photon
momentum $\bm{q}\sim 0$ here and hence there can only be vertical transitions in this diagram.
 We have chosen to show our results for $\theta=5^\circ$ as such an angle has been considered for other properties\cite{Choi:2011, Bistritzer:2011} and it is
intermediate to the data shown in Ref.~\cite{Brihuega:2012}. 
The conductivity curves are scaled by $\sigma_0=e^2/4\hbar$ which is the background conductivity of monolayer graphene.

The values of $\mu$ were chosen to span a range of behavior and to sample different
regions of the band structure. In Fig.~\ref{fig2}(d), charge neutrality is
considered with $\mu=0$. As expected, the low energy band structure seen in 
Fig.~\ref{fig2}(a) allows solely for interband transitions and a 
flat conductivity is found at low photon energy reflecting the low energy linear Dirac cones that are also found in monolayer graphene and emerge here with finite 
twist angle. The main difference is that there are now double the number of Dirac
cones compared to graphene due to the two layers in the bilayer, and hence the
universal background conductivity is $2\sigma_0$. This correlates with the
linear behavior in the low energy electronic density of states seen in 
STM and the monolayer behavior noted in those experiments. However, in the optics
the Fermi velocity does not enter this universal background value and as a result,
the reduction in $v_F$ with angle seen in STM experiments
would not be evident here. At higher photon frequency, for $\mu=0$, transitions
between the saddle point (VHS) in the band structure to other parts of the
band structure begin to occur. This results in the dip-peak structure seen
around $\Omega\sim 900$ meV, for the case shown here, which is very similar
to the structure calculated for unrotated Bernal-stacked bilayer graphene at charge
neutrality but found, in that case, at low energy starting from $\Omega=0$.\cite{Abergel:2007} 
Here, the peak arises from $k=0$ transitions involving the VHS and hence its observation should 
provide evidence for the VHS, even in the presence of reduced absorption at high energy due 
to limits on the $k$ integration that result from a Moir\'e Brillouin zone 
which is not included here. 
As $\theta$ is increased, this structure is moved to higher energy in 
our model, but the low energy behavior remains the same.

With doping away from charge neutrality to $\mu=240$ meV, the low energy interband
transitions are blocked by Pauli exclusion principle and intraband transitions,
facilitated by the impurity scattering rate, now give rise
to a narrow Drude absorption centered about $\Omega=0$. This Drude absorption
acquires the optical spectral weight that is removed at low frequencies
below $\Omega = 2\mu= 480$ meV. The dip-peak feature at higher photon energy remains the
same which would not be the case for ordinary Bernal-stacked bilayer graphene\cite{Nicol:2008}
where the peak would be split into two and reduced.
Thus, we find for low doping that the low energy behavior in the dynamical
conductivity will mimic the classic monolayer graphene behavior subject to
a factor of two increase in the overall magnitude of the conductivity.

Turning now to a more interesting result, shown in Fig.~\ref{fig2}(f),
if $\mu=450$ meV, the Fermi level is now above the low energy
saddle point (VHS) but below the next band in Fig.~\ref{fig2}(c). 
At this doping, a new peak appears in the conductivity at low energy marking the transition between the VHS to the second band,
which in this model measures $\tilde t_\perp$. As this is a result from transitions at $k=0$ and is at low photon
energy, it should be a very robust feature in far infrared measurements. 
Furthermore, the peak shown in the
model here is very strong. In graphene systems, the impurity scattering rate is
typically small on the order of a meV and hence, the width of the Drude 
would not be expected to interfere with this feature for intermediate 
rotation angle where our model applies. 
The spectral weight for this new VHS peak comes from the higher photon energy 
region in
the conductivity where a number of transitions involving the VHS point in
the band structure are now blocked. These are illustrated and understood by examining
the arrows shown in Fig.~\ref{fig2}(b) and (c). One sees that the dip-peak structure at
high energy is indeed diminished in this case. Thus, we predict that there 
should be a signature of the existence of the controversial low energy VHS in the
optical conductivity at low photon energy with appropriate doping which could be achieved
by voltage gating or by other means as has been done for ordinary bilayer graphene.\cite{Kuzmenko:2009a,Ohta:2006}

Finally, not shown here, but if $\mu$ is further increased to a value
which places it in the upper band ($\mu$ greater than $\sim 550$ meV),
then the VHS peak at low energy will  be lost due to Pauli blocking. Varying
$\mu$, therefore, would provide a sensitive probe as to the energy of
the saddle point VHS ($\varepsilon_{\rm VHS}\equiv\varepsilon_1(k=0)$) and the energy of the second
band at the $k=0$ point just above the saddle point, ie., $\varepsilon_2(k=0)$.
The photon energy for the absorption peak seen in Fig.~\ref{fig2}(f) would
give $|\varepsilon_2(k=0)-\varepsilon_{\rm VHS}|$ and the values of $\mu$ at the
first appearance and then disappearance of the peak would give 
$\varepsilon_{\rm VHS}$ and $\varepsilon_2(k=0)$, respectively, for a further
check on the numbers.

In summary, we have examined the optical conductivity of twisted bilayer graphene
using a simplified model for the low energy band structure in order to bring
out the qualitatively new features for this system. The monolayer graphene-like
 behavior
at low energies that has been seen in STM measurements manifests itself in the
conductivity as a flat universal background as a function of photon energy
but with magnitude twice that of the monolayer. Finite doping introduces Pauli blocking 
in the band structure for interband transitions but spectral weight is transferred to a Drude response due to intraband processes. At higher energy, features 
analogous to unrotated bilayer graphene may be found but shifted in energy and different in origin. The saddle point between
split Dirac cones in the band structure which is attributed to a VHS in the density of states, seen in STM, may be identified by the high energy
peak and the appearance
of a new peak at low energy in the dynamical conductivity when the Fermi level
sweeps through this region of the band structure. Thus, we provide a suggestion
for a way to confirm the existence of this VHS through the spectroscopy of
optical conductivity measurements which may provide further
confirmation of the existence of this unique band structure in twisted
bilayer graphene that has been a source of dispute.


We thank J.P. Carbotte and P. San-Jose for helpful discussions.
EJN acknowledges the hospitality of the KITP, Santa Barbara.
This work has been supported by NSERC of Canada
and in part by the National Science Foundation
 under Grant No. NSF PHY11-25915.  


\bibliographystyle{apsrev4-1}
\bibliography{twbi-bib}

\begin{thebibliography}{43}%
\makeatletter
\providecommand \@ifxundefined [1]{%
 \@ifx{#1\undefined}
}%
\providecommand \@ifnum [1]{%
 \ifnum #1\expandafter \@firstoftwo
 \else \expandafter \@secondoftwo
 \fi
}%
\providecommand \@ifx [1]{%
 \ifx #1\expandafter \@firstoftwo
 \else \expandafter \@secondoftwo
 \fi
}%
\providecommand \natexlab [1]{#1}%
\providecommand \enquote  [1]{``#1''}%
\providecommand \bibnamefont  [1]{#1}%
\providecommand \bibfnamefont [1]{#1}%
\providecommand \citenamefont [1]{#1}%
\providecommand \href@noop [0]{\@secondoftwo}%
\providecommand \href [0]{\begingroup \@sanitize@url \@href}%
\providecommand \@href[1]{\@@startlink{#1}\@@href}%
\providecommand \@@href[1]{\endgroup#1\@@endlink}%
\providecommand \@sanitize@url [0]{\catcode `\\12\catcode `\$12\catcode
  `\&12\catcode `\#12\catcode `\^12\catcode `\_12\catcode `\%12\relax}%
\providecommand \@@startlink[1]{}%
\providecommand \@@endlink[0]{}%
\providecommand \url  [0]{\begingroup\@sanitize@url \@url }%
\providecommand \@url [1]{\endgroup\@href {#1}{\urlprefix }}%
\providecommand \urlprefix  [0]{URL }%
\providecommand \Eprint [0]{\href }%
\providecommand \doibase [0]{http://dx.doi.org/}%
\providecommand \selectlanguage [0]{\@gobble}%
\providecommand \bibinfo  [0]{\@secondoftwo}%
\providecommand \bibfield  [0]{\@secondoftwo}%
\providecommand \translation [1]{[#1]}%
\providecommand \BibitemOpen [0]{}%
\providecommand \bibitemStop [0]{}%
\providecommand \bibitemNoStop [0]{.\EOS\space}%
\providecommand \EOS [0]{\spacefactor3000\relax}%
\providecommand \BibitemShut  [1]{\csname bibitem#1\endcsname}%
\let\auto@bib@innerbib\@empty
\bibitem [{\citenamefont {{Castro Neto}}\ \emph {et~al.}(2009)\citenamefont
  {{Castro Neto}}, \citenamefont {Guinea}, \citenamefont {Peres}, \citenamefont
  {Novoselov},\ and\ \citenamefont {Geim}}]{Neto:2009}%
  \BibitemOpen
  \bibfield  {author} {\bibinfo {author} {\bibfnamefont {A.~H.}\ \bibnamefont
  {{Castro Neto}}}, \bibinfo {author} {\bibfnamefont {F.}~\bibnamefont
  {Guinea}}, \bibinfo {author} {\bibfnamefont {N.~M.~R.}\ \bibnamefont
  {Peres}}, \bibinfo {author} {\bibfnamefont {K.~S.}\ \bibnamefont
  {Novoselov}}, \ and\ \bibinfo {author} {\bibfnamefont {A.~K.}\ \bibnamefont
  {Geim}},\ }\href@noop {} {\bibfield  {journal} {\bibinfo  {journal} {Rev.
  Mod. Phys.}\ }\textbf {\bibinfo {volume} {81}},\ \bibinfo {pages} {109}
  (\bibinfo {year} {2009})}\BibitemShut {NoStop}%
\bibitem [{\citenamefont {Abergel}\ \emph {et~al.}(2010)\citenamefont
  {Abergel}, \citenamefont {Apalkov}, \citenamefont {Berashevich},
  \citenamefont {Ziegler},\ and\ \citenamefont {Chakraborty}}]{Abergel:2010}%
  \BibitemOpen
  \bibfield  {author} {\bibinfo {author} {\bibfnamefont {D.~S.~L.}\
  \bibnamefont {Abergel}}, \bibinfo {author} {\bibfnamefont {V.}~\bibnamefont
  {Apalkov}}, \bibinfo {author} {\bibfnamefont {J.}~\bibnamefont
  {Berashevich}}, \bibinfo {author} {\bibfnamefont {K.}~\bibnamefont
  {Ziegler}}, \ and\ \bibinfo {author} {\bibfnamefont {T.}~\bibnamefont
  {Chakraborty}},\ }\href@noop {} {\bibfield  {journal} {\bibinfo  {journal}
  {Adv. in Phys.}\ }\textbf {\bibinfo {volume} {59}},\ \bibinfo {pages} {261}
  (\bibinfo {year} {2010})}\BibitemShut {NoStop}%
\bibitem [{\citenamefont {{Lopes dos Santos}}\ \emph
  {et~al.}(2007)\citenamefont {{Lopes dos Santos}}, \citenamefont {Peres},\
  and\ \citenamefont {{Castro Neto}}}]{Santos:2007}%
  \BibitemOpen
  \bibfield  {author} {\bibinfo {author} {\bibfnamefont {J.~M.~B.}\
  \bibnamefont {{Lopes dos Santos}}}, \bibinfo {author} {\bibfnamefont
  {N.~M.~R.}\ \bibnamefont {Peres}}, \ and\ \bibinfo {author} {\bibfnamefont
  {A.~H.}\ \bibnamefont {{Castro Neto}}},\ }\href@noop {} {\bibfield  {journal}
  {\bibinfo  {journal} {Phys. Rev. Lett.}\ }\textbf {\bibinfo {volume} {99}},\
  \bibinfo {pages} {256802} (\bibinfo {year} {2007})}\BibitemShut {NoStop}%
\bibitem [{\citenamefont {Li}\ \emph {et~al.}(2009{\natexlab{a}})\citenamefont
  {Li}, \citenamefont {Luican}, \citenamefont {dos Santos}, \citenamefont
  {Neto}, \citenamefont {Reina}, \citenamefont {Kong},\ and\ \citenamefont
  {Andrei}}]{Li:2009}%
  \BibitemOpen
  \bibfield  {author} {\bibinfo {author} {\bibfnamefont {G.}~\bibnamefont
  {Li}}, \bibinfo {author} {\bibfnamefont {A.}~\bibnamefont {Luican}}, \bibinfo
  {author} {\bibfnamefont {J.~L.}\ \bibnamefont {dos Santos}}, \bibinfo
  {author} {\bibfnamefont {A.~C.}\ \bibnamefont {Neto}}, \bibinfo {author}
  {\bibfnamefont {A.}~\bibnamefont {Reina}}, \bibinfo {author} {\bibfnamefont
  {J.}~\bibnamefont {Kong}}, \ and\ \bibinfo {author} {\bibfnamefont
  {E.}~\bibnamefont {Andrei}},\ }\href@noop {} {\bibfield  {journal} {\bibinfo
  {journal} {Nature Phys.}\ }\textbf {\bibinfo {volume} {6}},\ \bibinfo {pages}
  {109} (\bibinfo {year} {2009}{\natexlab{a}})}\BibitemShut {NoStop}%
\bibitem [{\citenamefont {Luican}\ \emph {et~al.}(2011)\citenamefont {Luican},
  \citenamefont {Li}, \citenamefont {Reina}, \citenamefont {Kong},
  \citenamefont {Nair}, \citenamefont {Novoselov}, \citenamefont {Geim},\ and\
  \citenamefont {Andrei}}]{Luican:2011}%
  \BibitemOpen
  \bibfield  {author} {\bibinfo {author} {\bibfnamefont {A.}~\bibnamefont
  {Luican}}, \bibinfo {author} {\bibfnamefont {G.}~\bibnamefont {Li}}, \bibinfo
  {author} {\bibfnamefont {A.}~\bibnamefont {Reina}}, \bibinfo {author}
  {\bibfnamefont {J.}~\bibnamefont {Kong}}, \bibinfo {author} {\bibfnamefont
  {R.~R.}\ \bibnamefont {Nair}}, \bibinfo {author} {\bibfnamefont {K.~S.}\
  \bibnamefont {Novoselov}}, \bibinfo {author} {\bibfnamefont {A.~K.}\
  \bibnamefont {Geim}}, \ and\ \bibinfo {author} {\bibfnamefont {E.~Y.}\
  \bibnamefont {Andrei}},\ }\href {\doibase 10.1103/PhysRevLett.106.126802}
  {\bibfield  {journal} {\bibinfo  {journal} {Phys. Rev. Lett.}\ }\textbf
  {\bibinfo {volume} {106}},\ \bibinfo {pages} {126802} (\bibinfo {year}
  {2011})}\BibitemShut {NoStop}%
\bibitem [{\citenamefont {Brihuega}\ \emph {et~al.}(2012)\citenamefont
  {Brihuega}, \citenamefont {Mallet}, \citenamefont {Gonzalez-Herrero},
  \citenamefont {de~Laissardiere}, \citenamefont {Ugeda}, \citenamefont
  {Magaud}, \citenamefont {Gomez-Rodriguez}, \citenamefont {Yndurain},\ and\
  \citenamefont {Veuillen}}]{Brihuega:2012}%
  \BibitemOpen
  \bibfield  {author} {\bibinfo {author} {\bibfnamefont {I.}~\bibnamefont
  {Brihuega}}, \bibinfo {author} {\bibfnamefont {P.}~\bibnamefont {Mallet}},
  \bibinfo {author} {\bibfnamefont {H.}~\bibnamefont {Gonzalez-Herrero}},
  \bibinfo {author} {\bibfnamefont {G.~T.}\ \bibnamefont {de~Laissardiere}},
  \bibinfo {author} {\bibfnamefont {M.}~\bibnamefont {Ugeda}}, \bibinfo
  {author} {\bibfnamefont {L.}~\bibnamefont {Magaud}}, \bibinfo {author}
  {\bibfnamefont {J.}~\bibnamefont {Gomez-Rodriguez}}, \bibinfo {author}
  {\bibfnamefont {F.}~\bibnamefont {Yndurain}}, \ and\ \bibinfo {author}
  {\bibfnamefont {J.-Y.}\ \bibnamefont {Veuillen}},\ }\href@noop {} {\bibfield
  {journal} {\bibinfo  {journal} {Phys. Rev. Lett.}\ }\textbf {\bibinfo
  {volume} {109}},\ \bibinfo {pages} {196802} (\bibinfo {year}
  {2012})}\BibitemShut {NoStop}%
\bibitem [{\citenamefont {Trambly~de Laissardi\`ere}\ \emph
  {et~al.}(2010)\citenamefont {Trambly~de Laissardi\`ere}, \citenamefont
  {Mayou},\ and\ \citenamefont {Magaud}}]{Trambly:2010}%
  \BibitemOpen
  \bibfield  {author} {\bibinfo {author} {\bibfnamefont {G.}~\bibnamefont
  {Trambly~de Laissardi\`ere}}, \bibinfo {author} {\bibfnamefont
  {D.}~\bibnamefont {Mayou}}, \ and\ \bibinfo {author} {\bibfnamefont
  {L.}~\bibnamefont {Magaud}},\ }\href@noop {} {\bibfield  {journal} {\bibinfo
  {journal} {Nano Lett.}\ }\textbf {\bibinfo {volume} {10}},\ \bibinfo {pages}
  {804} (\bibinfo {year} {2010})}\BibitemShut {NoStop}%
\bibitem [{\citenamefont {San-Jose}\ \emph {et~al.}(2012)\citenamefont
  {San-Jose}, \citenamefont {Gonz\'alez},\ and\ \citenamefont
  {Guinea}}]{San-Jose:2012}%
  \BibitemOpen
  \bibfield  {author} {\bibinfo {author} {\bibfnamefont {P.}~\bibnamefont
  {San-Jose}}, \bibinfo {author} {\bibfnamefont {J.}~\bibnamefont
  {Gonz\'alez}}, \ and\ \bibinfo {author} {\bibfnamefont {F.}~\bibnamefont
  {Guinea}},\ }\href {\doibase 10.1103/PhysRevLett.108.216802} {\bibfield
  {journal} {\bibinfo  {journal} {Phys. Rev. Lett.}\ }\textbf {\bibinfo
  {volume} {108}},\ \bibinfo {pages} {216802} (\bibinfo {year}
  {2012})}\BibitemShut {NoStop}%
\bibitem [{\citenamefont {Bistritzer}\ and\ \citenamefont
  {MacDonald}(2011)}]{Bistritzer:2011}%
  \BibitemOpen
  \bibfield  {author} {\bibinfo {author} {\bibfnamefont {R.}~\bibnamefont
  {Bistritzer}}\ and\ \bibinfo {author} {\bibfnamefont {A.~H.}\ \bibnamefont
  {MacDonald}},\ }\href {\doibase doi:10.1073/pnas.1108174108} {\bibfield
  {journal} {\bibinfo  {journal} {Proc. Nat. Acad. Sci. USA}\ }\textbf
  {\bibinfo {volume} {108}},\ \bibinfo {pages} {12233} (\bibinfo {year}
  {2011})}\BibitemShut {NoStop}%
\bibitem [{\citenamefont {Hicks}\ \emph {et~al.}(2011)\citenamefont {Hicks},
  \citenamefont {Sprinkle}, \citenamefont {Shepperd}, \citenamefont {Wang},
  \citenamefont {Tejeda}, \citenamefont {Taleb-Ibrahimi}, \citenamefont
  {Bertran}, \citenamefont {Le~F\`evre}, \citenamefont {de~Heer}, \citenamefont
  {Berger},\ and\ \citenamefont {Conrad}}]{Hicks:2011}%
  \BibitemOpen
  \bibfield  {author} {\bibinfo {author} {\bibfnamefont {J.}~\bibnamefont
  {Hicks}}, \bibinfo {author} {\bibfnamefont {M.}~\bibnamefont {Sprinkle}},
  \bibinfo {author} {\bibfnamefont {K.}~\bibnamefont {Shepperd}}, \bibinfo
  {author} {\bibfnamefont {F.}~\bibnamefont {Wang}}, \bibinfo {author}
  {\bibfnamefont {A.}~\bibnamefont {Tejeda}}, \bibinfo {author} {\bibfnamefont
  {A.}~\bibnamefont {Taleb-Ibrahimi}}, \bibinfo {author} {\bibfnamefont
  {F.}~\bibnamefont {Bertran}}, \bibinfo {author} {\bibfnamefont
  {P.}~\bibnamefont {Le~F\`evre}}, \bibinfo {author} {\bibfnamefont {W.~A.}\
  \bibnamefont {de~Heer}}, \bibinfo {author} {\bibfnamefont {C.}~\bibnamefont
  {Berger}}, \ and\ \bibinfo {author} {\bibfnamefont {E.~H.}\ \bibnamefont
  {Conrad}},\ }\href {\doibase 10.1103/PhysRevB.83.205403} {\bibfield
  {journal} {\bibinfo  {journal} {Phys. Rev. B}\ }\textbf {\bibinfo {volume}
  {83}},\ \bibinfo {pages} {205403} (\bibinfo {year} {2011})}\BibitemShut
  {NoStop}%
\bibitem [{\citenamefont {Ohta}\ \emph {et~al.}(2012)\citenamefont {Ohta},
  \citenamefont {Robinson}, \citenamefont {Feibelman}, \citenamefont
  {Bostwick}, \citenamefont {Rotenberg},\ and\ \citenamefont
  {Beechem}}]{Ohta:2012}%
  \BibitemOpen
  \bibfield  {author} {\bibinfo {author} {\bibfnamefont {T.}~\bibnamefont
  {Ohta}}, \bibinfo {author} {\bibfnamefont {J.~T.}\ \bibnamefont {Robinson}},
  \bibinfo {author} {\bibfnamefont {P.~J.}\ \bibnamefont {Feibelman}}, \bibinfo
  {author} {\bibfnamefont {A.}~\bibnamefont {Bostwick}}, \bibinfo {author}
  {\bibfnamefont {E.}~\bibnamefont {Rotenberg}}, \ and\ \bibinfo {author}
  {\bibfnamefont {T.~E.}\ \bibnamefont {Beechem}},\ }\href {\doibase
  10.1103/PhysRevLett.109.186807} {\bibfield  {journal} {\bibinfo  {journal}
  {Phys. Rev. Lett.}\ }\textbf {\bibinfo {volume} {109}},\ \bibinfo {pages}
  {186807} (\bibinfo {year} {2012})}\BibitemShut {NoStop}%
\bibitem [{\citenamefont {{de Gail}}\ \emph {et~al.}(2011)\citenamefont {{de
  Gail}}, \citenamefont {Goerbig}, \citenamefont {Guinea}, \citenamefont
  {Montambaux},\ and\ \citenamefont {{Castro Neto}}}]{deGail:2011}%
  \BibitemOpen
  \bibfield  {author} {\bibinfo {author} {\bibfnamefont {R.}~\bibnamefont {{de
  Gail}}}, \bibinfo {author} {\bibfnamefont {M.~O.}\ \bibnamefont {Goerbig}},
  \bibinfo {author} {\bibfnamefont {F.}~\bibnamefont {Guinea}}, \bibinfo
  {author} {\bibfnamefont {G.}~\bibnamefont {Montambaux}}, \ and\ \bibinfo
  {author} {\bibfnamefont {A.~H.}\ \bibnamefont {{Castro Neto}}},\ }\href@noop
  {} {\bibfield  {journal} {\bibinfo  {journal} {Phys. Rev. B}\ }\textbf
  {\bibinfo {volume} {84}},\ \bibinfo {pages} {045436} (\bibinfo {year}
  {2011})}\BibitemShut {NoStop}%
\bibitem [{\citenamefont {Choi}\ \emph {et~al.}(2011)\citenamefont {Choi},
  \citenamefont {Hyun},\ and\ \citenamefont {Kim}}]{Choi:2011}%
  \BibitemOpen
  \bibfield  {author} {\bibinfo {author} {\bibfnamefont {M.-Y.}\ \bibnamefont
  {Choi}}, \bibinfo {author} {\bibfnamefont {Y.-H.}\ \bibnamefont {Hyun}}, \
  and\ \bibinfo {author} {\bibfnamefont {Y.}~\bibnamefont {Kim}},\ }\href@noop
  {} {\bibfield  {journal} {\bibinfo  {journal} {Phys. Rev. B}\ }\textbf
  {\bibinfo {volume} {84}},\ \bibinfo {pages} {195437} (\bibinfo {year}
  {2011})}\BibitemShut {NoStop}%
\bibitem [{\citenamefont {Lee}\ \emph {et~al.}(2011)\citenamefont {Lee},
  \citenamefont {Riedl}, \citenamefont {Beringer}, \citenamefont {Castro~Neto},
  \citenamefont {von Klitzing}, \citenamefont {Starke},\ and\ \citenamefont
  {Smet}}]{Lee:2011}%
  \BibitemOpen
  \bibfield  {author} {\bibinfo {author} {\bibfnamefont {D.~S.}\ \bibnamefont
  {Lee}}, \bibinfo {author} {\bibfnamefont {C.}~\bibnamefont {Riedl}}, \bibinfo
  {author} {\bibfnamefont {T.}~\bibnamefont {Beringer}}, \bibinfo {author}
  {\bibfnamefont {A.~H.}\ \bibnamefont {Castro~Neto}}, \bibinfo {author}
  {\bibfnamefont {K.}~\bibnamefont {von Klitzing}}, \bibinfo {author}
  {\bibfnamefont {U.}~\bibnamefont {Starke}}, \ and\ \bibinfo {author}
  {\bibfnamefont {J.~H.}\ \bibnamefont {Smet}},\ }\href {\doibase
  10.1103/PhysRevLett.107.216602} {\bibfield  {journal} {\bibinfo  {journal}
  {Phys. Rev. Lett.}\ }\textbf {\bibinfo {volume} {107}},\ \bibinfo {pages}
  {216602} (\bibinfo {year} {2011})}\BibitemShut {NoStop}%
\bibitem [{\citenamefont {Kim}\ \emph {et~al.}(2012)\citenamefont {Kim},
  \citenamefont {Coh}, \citenamefont {Tan}, \citenamefont {Regan},
  \citenamefont {Yuk}, \citenamefont {Chatterjee}, \citenamefont {Crommie},
  \citenamefont {Cohen}, \citenamefont {Louie},\ and\ \citenamefont
  {Zettl}}]{Kim:2012}%
  \BibitemOpen
  \bibfield  {author} {\bibinfo {author} {\bibfnamefont {K.}~\bibnamefont
  {Kim}}, \bibinfo {author} {\bibfnamefont {S.}~\bibnamefont {Coh}}, \bibinfo
  {author} {\bibfnamefont {L.~Z.}\ \bibnamefont {Tan}}, \bibinfo {author}
  {\bibfnamefont {W.}~\bibnamefont {Regan}}, \bibinfo {author} {\bibfnamefont
  {J.~M.}\ \bibnamefont {Yuk}}, \bibinfo {author} {\bibfnamefont
  {E.}~\bibnamefont {Chatterjee}}, \bibinfo {author} {\bibfnamefont {M.~F.}\
  \bibnamefont {Crommie}}, \bibinfo {author} {\bibfnamefont {M.~L.}\
  \bibnamefont {Cohen}}, \bibinfo {author} {\bibfnamefont {S.~G.}\ \bibnamefont
  {Louie}}, \ and\ \bibinfo {author} {\bibfnamefont {A.}~\bibnamefont
  {Zettl}},\ }\href {\doibase 10.1103/PhysRevLett.108.246103} {\bibfield
  {journal} {\bibinfo  {journal} {Phys. Rev. Lett.}\ }\textbf {\bibinfo
  {volume} {108}},\ \bibinfo {pages} {246103} (\bibinfo {year}
  {2012})}\BibitemShut {NoStop}%
\bibitem [{\citenamefont {Ando}\ \emph {et~al.}(2002)\citenamefont {Ando},
  \citenamefont {Zheng},\ and\ \citenamefont {Suzuura}}]{Ando:2002}%
  \BibitemOpen
  \bibfield  {author} {\bibinfo {author} {\bibfnamefont {T.}~\bibnamefont
  {Ando}}, \bibinfo {author} {\bibfnamefont {Y.}~\bibnamefont {Zheng}}, \ and\
  \bibinfo {author} {\bibfnamefont {H.}~\bibnamefont {Suzuura}},\ }\href@noop
  {} {\bibfield  {journal} {\bibinfo  {journal} {J. Phys. Soc. Jpn.}\ }\textbf
  {\bibinfo {volume} {71}},\ \bibinfo {pages} {1318} (\bibinfo {year}
  {2002})}\BibitemShut {NoStop}%
\bibitem [{\citenamefont {Gusynin}\ and\ \citenamefont
  {Sharapov}(2006)}]{Gusynin:2006a}%
  \BibitemOpen
  \bibfield  {author} {\bibinfo {author} {\bibfnamefont {V.~P.}\ \bibnamefont
  {Gusynin}}\ and\ \bibinfo {author} {\bibfnamefont {S.~G.}\ \bibnamefont
  {Sharapov}},\ }\href@noop {} {\bibfield  {journal} {\bibinfo  {journal}
  {Phys. Rev. B}\ }\textbf {\bibinfo {volume} {73}},\ \bibinfo {pages} {245411}
  (\bibinfo {year} {2006})}\BibitemShut {NoStop}%
\bibitem [{\citenamefont {Gusynin}\ \emph {et~al.}(2006)\citenamefont
  {Gusynin}, \citenamefont {Sharapov},\ and\ \citenamefont
  {Carbotte}}]{Gusynin:2006}%
  \BibitemOpen
  \bibfield  {author} {\bibinfo {author} {\bibfnamefont {V.~P.}\ \bibnamefont
  {Gusynin}}, \bibinfo {author} {\bibfnamefont {S.~G.}\ \bibnamefont
  {Sharapov}}, \ and\ \bibinfo {author} {\bibfnamefont {J.~P.}\ \bibnamefont
  {Carbotte}},\ }\href@noop {} {\bibfield  {journal} {\bibinfo  {journal}
  {Phys. Rev. Lett.}\ }\textbf {\bibinfo {volume} {96}},\ \bibinfo {pages}
  {256802} (\bibinfo {year} {2006})}\BibitemShut {NoStop}%
\bibitem [{\citenamefont {Peres}\ \emph {et~al.}(2006)\citenamefont {Peres},
  \citenamefont {Guinea},\ and\ \citenamefont {Castro~Neto}}]{Peres:2006}%
  \BibitemOpen
  \bibfield  {author} {\bibinfo {author} {\bibfnamefont {N.~M.~R.}\
  \bibnamefont {Peres}}, \bibinfo {author} {\bibfnamefont {F.}~\bibnamefont
  {Guinea}}, \ and\ \bibinfo {author} {\bibfnamefont {A.~H.}\ \bibnamefont
  {Castro~Neto}},\ }\href {\doibase 10.1103/PhysRevB.73.125411} {\bibfield
  {journal} {\bibinfo  {journal} {Phys. Rev. B}\ }\textbf {\bibinfo {volume}
  {73}},\ \bibinfo {pages} {125411} (\bibinfo {year} {2006})}\BibitemShut
  {NoStop}%
\bibitem [{\citenamefont {Falkovsky}\ and\ \citenamefont
  {Varlamov}(2007)}]{Falkovsky:2007}%
  \BibitemOpen
  \bibfield  {author} {\bibinfo {author} {\bibfnamefont {L.~A.}\ \bibnamefont
  {Falkovsky}}\ and\ \bibinfo {author} {\bibfnamefont {A.~A.}\ \bibnamefont
  {Varlamov}},\ }\href@noop {} {\bibfield  {journal} {\bibinfo  {journal} {Eur.
  Phys. J. B}\ }\textbf {\bibinfo {volume} {56}},\ \bibinfo {pages} {281}
  (\bibinfo {year} {2007})}\BibitemShut {NoStop}%
\bibitem [{\citenamefont {Stauber}\ \emph {et~al.}(2008)\citenamefont
  {Stauber}, \citenamefont {Peres},\ and\ \citenamefont {{Castro
  Neto}}}]{Stauber:2008}%
  \BibitemOpen
  \bibfield  {author} {\bibinfo {author} {\bibfnamefont {T.}~\bibnamefont
  {Stauber}}, \bibinfo {author} {\bibfnamefont {N.~M.~R.}\ \bibnamefont
  {Peres}}, \ and\ \bibinfo {author} {\bibfnamefont {A.~H.}\ \bibnamefont
  {{Castro Neto}}},\ }\href@noop {} {\bibfield  {journal} {\bibinfo  {journal}
  {Phys. Rev. B}\ }\textbf {\bibinfo {volume} {78}},\ \bibinfo {pages} {085418}
  (\bibinfo {year} {2008})}\BibitemShut {NoStop}%
\bibitem [{\citenamefont {Abergel}\ and\ \citenamefont
  {Fal'ko}(2007)}]{Abergel:2007}%
  \BibitemOpen
  \bibfield  {author} {\bibinfo {author} {\bibfnamefont {D.~S.~L.}\
  \bibnamefont {Abergel}}\ and\ \bibinfo {author} {\bibfnamefont {V.~I.}\
  \bibnamefont {Fal'ko}},\ }\href {\doibase 10.1103/PhysRevB.75.155430}
  {\bibfield  {journal} {\bibinfo  {journal} {Phys. Rev. B}\ }\textbf {\bibinfo
  {volume} {75}},\ \bibinfo {pages} {155430} (\bibinfo {year}
  {2007})}\BibitemShut {NoStop}%
\bibitem [{\citenamefont {Nilsson}\ \emph {et~al.}(2008)\citenamefont
  {Nilsson}, \citenamefont {{Castro Neto}}, \citenamefont {Guinea},\ and\
  \citenamefont {Peres}}]{Nilsson:2008}%
  \BibitemOpen
  \bibfield  {author} {\bibinfo {author} {\bibfnamefont {J.}~\bibnamefont
  {Nilsson}}, \bibinfo {author} {\bibfnamefont {A.~H.}\ \bibnamefont {{Castro
  Neto}}}, \bibinfo {author} {\bibfnamefont {F.}~\bibnamefont {Guinea}}, \ and\
  \bibinfo {author} {\bibfnamefont {N.~M.~R.}\ \bibnamefont {Peres}},\
  }\href@noop {} {\bibfield  {journal} {\bibinfo  {journal} {Phys. Rev. B}\
  }\textbf {\bibinfo {volume} {78}},\ \bibinfo {pages} {045405} (\bibinfo
  {year} {2008})}\BibitemShut {NoStop}%
\bibitem [{\citenamefont {Nicol}\ and\ \citenamefont
  {Carbotte}(2008)}]{Nicol:2008}%
  \BibitemOpen
  \bibfield  {author} {\bibinfo {author} {\bibfnamefont {E.~J.}\ \bibnamefont
  {Nicol}}\ and\ \bibinfo {author} {\bibfnamefont {J.~P.}\ \bibnamefont
  {Carbotte}},\ }\href@noop {} {\bibfield  {journal} {\bibinfo  {journal}
  {Phys. Rev. B}\ }\textbf {\bibinfo {volume} {77}},\ \bibinfo {pages} {155409}
  (\bibinfo {year} {2008})}\BibitemShut {NoStop}%
\bibitem [{\citenamefont {Koshino}\ and\ \citenamefont
  {Ando}(2009)}]{Koshino:2009}%
  \BibitemOpen
  \bibfield  {author} {\bibinfo {author} {\bibfnamefont {M.}~\bibnamefont
  {Koshino}}\ and\ \bibinfo {author} {\bibfnamefont {T.}~\bibnamefont {Ando}},\
  }\href@noop {} {\bibfield  {journal} {\bibinfo  {journal} {Solid State
  Commun.}\ }\textbf {\bibinfo {volume} {149}},\ \bibinfo {pages} {1123}
  (\bibinfo {year} {2009})}\BibitemShut {NoStop}%
\bibitem [{\citenamefont {Zhang}\ \emph {et~al.}(2008)\citenamefont {Zhang},
  \citenamefont {Li}, \citenamefont {Basov}, \citenamefont {Fogler},
  \citenamefont {Hao},\ and\ \citenamefont {Martin}}]{Zhang:2008}%
  \BibitemOpen
  \bibfield  {author} {\bibinfo {author} {\bibfnamefont {L.~M.}\ \bibnamefont
  {Zhang}}, \bibinfo {author} {\bibfnamefont {Z.~Q.}\ \bibnamefont {Li}},
  \bibinfo {author} {\bibfnamefont {D.~N.}\ \bibnamefont {Basov}}, \bibinfo
  {author} {\bibfnamefont {M.~M.}\ \bibnamefont {Fogler}}, \bibinfo {author}
  {\bibfnamefont {Z.}~\bibnamefont {Hao}}, \ and\ \bibinfo {author}
  {\bibfnamefont {M.~C.}\ \bibnamefont {Martin}},\ }\href@noop {} {\bibfield
  {journal} {\bibinfo  {journal} {Phys. Rev. B}\ }\textbf {\bibinfo {volume}
  {78}},\ \bibinfo {pages} {235408} (\bibinfo {year} {2008})}\BibitemShut
  {NoStop}%
\bibitem [{\citenamefont {Li}\ \emph {et~al.}(2008)\citenamefont {Li},
  \citenamefont {Henriksen}, \citenamefont {Jiang}, \citenamefont {Hao},
  \citenamefont {Martin}, \citenamefont {Kim}, \citenamefont {Stormer},\ and\
  \citenamefont {Basov}}]{Li:2008}%
  \BibitemOpen
  \bibfield  {author} {\bibinfo {author} {\bibfnamefont {Z.~Q.}\ \bibnamefont
  {Li}}, \bibinfo {author} {\bibfnamefont {E.~A.}\ \bibnamefont {Henriksen}},
  \bibinfo {author} {\bibfnamefont {Z.}~\bibnamefont {Jiang}}, \bibinfo
  {author} {\bibfnamefont {Z.}~\bibnamefont {Hao}}, \bibinfo {author}
  {\bibfnamefont {M.~C.}\ \bibnamefont {Martin}}, \bibinfo {author}
  {\bibfnamefont {P.}~\bibnamefont {Kim}}, \bibinfo {author} {\bibfnamefont
  {H.~L.}\ \bibnamefont {Stormer}}, \ and\ \bibinfo {author} {\bibfnamefont
  {D.~N.}\ \bibnamefont {Basov}},\ }\href@noop {} {\bibfield  {journal}
  {\bibinfo  {journal} {Nature Phys.}\ }\textbf {\bibinfo {volume} {4}},\
  \bibinfo {pages} {532} (\bibinfo {year} {2008})}\BibitemShut {NoStop}%
\bibitem [{\citenamefont {Kuzmenko}\ \emph {et~al.}(2008)\citenamefont
  {Kuzmenko}, \citenamefont {van Heumen}, \citenamefont {Carbone},\ and\
  \citenamefont {van~der Marel}}]{Kuzmenko:2008}%
  \BibitemOpen
  \bibfield  {author} {\bibinfo {author} {\bibfnamefont {A.~B.}\ \bibnamefont
  {Kuzmenko}}, \bibinfo {author} {\bibfnamefont {E.}~\bibnamefont {van
  Heumen}}, \bibinfo {author} {\bibfnamefont {F.}~\bibnamefont {Carbone}}, \
  and\ \bibinfo {author} {\bibfnamefont {D.}~\bibnamefont {van~der Marel}},\
  }\href@noop {} {\bibfield  {journal} {\bibinfo  {journal} {Phys. Rev. Lett.}\
  }\textbf {\bibinfo {volume} {100}},\ \bibinfo {pages} {117401} (\bibinfo
  {year} {2008})}\BibitemShut {NoStop}%
\bibitem [{\citenamefont {Wang}\ \emph {et~al.}(2008)\citenamefont {Wang},
  \citenamefont {Zhang}, \citenamefont {Tian}, \citenamefont {Girit},
  \citenamefont {Zettl}, \citenamefont {Crommie},\ and\ \citenamefont
  {Shen}}]{Wang:2008}%
  \BibitemOpen
  \bibfield  {author} {\bibinfo {author} {\bibfnamefont {F.}~\bibnamefont
  {Wang}}, \bibinfo {author} {\bibfnamefont {Y.}~\bibnamefont {Zhang}},
  \bibinfo {author} {\bibfnamefont {C.}~\bibnamefont {Tian}}, \bibinfo {author}
  {\bibfnamefont {C.}~\bibnamefont {Girit}}, \bibinfo {author} {\bibfnamefont
  {A.}~\bibnamefont {Zettl}}, \bibinfo {author} {\bibfnamefont
  {M.}~\bibnamefont {Crommie}}, \ and\ \bibinfo {author} {\bibfnamefont
  {Y.}~\bibnamefont {Shen}},\ }\href@noop {} {\bibfield  {journal} {\bibinfo
  {journal} {Science}\ }\textbf {\bibinfo {volume} {320}},\ \bibinfo {pages}
  {206} (\bibinfo {year} {2008})}\BibitemShut {NoStop}%
\bibitem [{\citenamefont {Nair}\ \emph {et~al.}(2008)\citenamefont {Nair},
  \citenamefont {Blake}, \citenamefont {Grigeronko}, \citenamefont {Novoselov},
  \citenamefont {Booth}, \citenamefont {Stauber}, \citenamefont {Peres},\ and\
  \citenamefont {Geim}}]{Nair:2008}%
  \BibitemOpen
  \bibfield  {author} {\bibinfo {author} {\bibfnamefont {R.~R.}\ \bibnamefont
  {Nair}}, \bibinfo {author} {\bibfnamefont {B.}~\bibnamefont {Blake}},
  \bibinfo {author} {\bibfnamefont {A.~N.}\ \bibnamefont {Grigeronko}},
  \bibinfo {author} {\bibfnamefont {K.~S.}\ \bibnamefont {Novoselov}}, \bibinfo
  {author} {\bibfnamefont {T.~J.}\ \bibnamefont {Booth}}, \bibinfo {author}
  {\bibfnamefont {T.}~\bibnamefont {Stauber}}, \bibinfo {author} {\bibfnamefont
  {N.~M.~R.}\ \bibnamefont {Peres}}, \ and\ \bibinfo {author} {\bibfnamefont
  {A.~K.}\ \bibnamefont {Geim}},\ }\href@noop {} {\bibfield  {journal}
  {\bibinfo  {journal} {Science}\ }\textbf {\bibinfo {volume} {320}},\ \bibinfo
  {pages} {1308} (\bibinfo {year} {2008})}\BibitemShut {NoStop}%
\bibitem [{\citenamefont {Mak}\ \emph {et~al.}(2008)\citenamefont {Mak},
  \citenamefont {Sfeir}, \citenamefont {Wu}, \citenamefont {Lui}, \citenamefont
  {Misewich},\ and\ \citenamefont {Heinz}}]{Mak:2008}%
  \BibitemOpen
  \bibfield  {author} {\bibinfo {author} {\bibfnamefont {K.~F.}\ \bibnamefont
  {Mak}}, \bibinfo {author} {\bibfnamefont {M.~Y.}\ \bibnamefont {Sfeir}},
  \bibinfo {author} {\bibfnamefont {Y.}~\bibnamefont {Wu}}, \bibinfo {author}
  {\bibfnamefont {C.~H.}\ \bibnamefont {Lui}}, \bibinfo {author} {\bibfnamefont
  {J.~A.}\ \bibnamefont {Misewich}}, \ and\ \bibinfo {author} {\bibfnamefont
  {T.~F.}\ \bibnamefont {Heinz}},\ }\href@noop {} {\bibfield  {journal}
  {\bibinfo  {journal} {Phys. Rev. Lett.}\ }\textbf {\bibinfo {volume} {101}},\
  \bibinfo {pages} {196405} (\bibinfo {year} {2008})}\BibitemShut {NoStop}%
\bibitem [{\citenamefont {Li}\ \emph {et~al.}(2009{\natexlab{b}})\citenamefont
  {Li}, \citenamefont {Henriksen}, \citenamefont {Jiang}, \citenamefont {Hao},
  \citenamefont {Martin}, \citenamefont {Kim}, \citenamefont {Stormer},\ and\
  \citenamefont {Basov}}]{ZLi:2009}%
  \BibitemOpen
  \bibfield  {author} {\bibinfo {author} {\bibfnamefont {Z.~Q.}\ \bibnamefont
  {Li}}, \bibinfo {author} {\bibfnamefont {E.~A.}\ \bibnamefont {Henriksen}},
  \bibinfo {author} {\bibfnamefont {Z.}~\bibnamefont {Jiang}}, \bibinfo
  {author} {\bibfnamefont {Z.}~\bibnamefont {Hao}}, \bibinfo {author}
  {\bibfnamefont {M.~C.}\ \bibnamefont {Martin}}, \bibinfo {author}
  {\bibfnamefont {P.}~\bibnamefont {Kim}}, \bibinfo {author} {\bibfnamefont
  {H.~L.}\ \bibnamefont {Stormer}}, \ and\ \bibinfo {author} {\bibfnamefont
  {D.~N.}\ \bibnamefont {Basov}},\ }\href {\doibase
  10.1103/PhysRevLett.102.037403} {\bibfield  {journal} {\bibinfo  {journal}
  {Phys. Rev. Lett.}\ }\textbf {\bibinfo {volume} {102}},\ \bibinfo {pages}
  {037403} (\bibinfo {year} {2009}{\natexlab{b}})}\BibitemShut {NoStop}%
\bibitem [{\citenamefont {Kuzmenko}\ \emph
  {et~al.}(2009{\natexlab{a}})\citenamefont {Kuzmenko}, \citenamefont {van
  Heumen}, \citenamefont {van~der Marel}, \citenamefont {Lerch}, \citenamefont
  {Blake}, \citenamefont {Novoselov},\ and\ \citenamefont
  {Geim}}]{Kuzmenko:2009a}%
  \BibitemOpen
  \bibfield  {author} {\bibinfo {author} {\bibfnamefont {A.~B.}\ \bibnamefont
  {Kuzmenko}}, \bibinfo {author} {\bibfnamefont {E.}~\bibnamefont {van
  Heumen}}, \bibinfo {author} {\bibfnamefont {D.}~\bibnamefont {van~der
  Marel}}, \bibinfo {author} {\bibfnamefont {P.}~\bibnamefont {Lerch}},
  \bibinfo {author} {\bibfnamefont {P.}~\bibnamefont {Blake}}, \bibinfo
  {author} {\bibfnamefont {K.~S.}\ \bibnamefont {Novoselov}}, \ and\ \bibinfo
  {author} {\bibfnamefont {A.~K.}\ \bibnamefont {Geim}},\ }\href@noop {}
  {\bibfield  {journal} {\bibinfo  {journal} {Phys. Rev. B}\ }\textbf {\bibinfo
  {volume} {79}},\ \bibinfo {pages} {115441} (\bibinfo {year}
  {2009}{\natexlab{a}})}\BibitemShut {NoStop}%
\bibitem [{\citenamefont {Kuzmenko}\ \emph
  {et~al.}(2009{\natexlab{b}})\citenamefont {Kuzmenko}, \citenamefont
  {Crassee}, \citenamefont {van~der Marel}, \citenamefont {Blake},\ and\
  \citenamefont {Novoselov}}]{Kuzmenko:2009b}%
  \BibitemOpen
  \bibfield  {author} {\bibinfo {author} {\bibfnamefont {A.~B.}\ \bibnamefont
  {Kuzmenko}}, \bibinfo {author} {\bibfnamefont {I.}~\bibnamefont {Crassee}},
  \bibinfo {author} {\bibfnamefont {D.}~\bibnamefont {van~der Marel}}, \bibinfo
  {author} {\bibfnamefont {P.}~\bibnamefont {Blake}}, \ and\ \bibinfo {author}
  {\bibfnamefont {K.~S.}\ \bibnamefont {Novoselov}},\ }\href {\doibase
  10.1103/PhysRevB.80.165406} {\bibfield  {journal} {\bibinfo  {journal} {Phys.
  Rev. B}\ }\textbf {\bibinfo {volume} {80}},\ \bibinfo {pages} {165406}
  (\bibinfo {year} {2009}{\natexlab{b}})}\BibitemShut {NoStop}%
\bibitem [{\citenamefont {Orlita}\ and\ \citenamefont
  {Potemski}(2010)}]{Orlita:2010}%
  \BibitemOpen
  \bibfield  {author} {\bibinfo {author} {\bibfnamefont {M.}~\bibnamefont
  {Orlita}}\ and\ \bibinfo {author} {\bibfnamefont {M.}~\bibnamefont
  {Potemski}},\ }\href@noop {} {\bibfield  {journal} {\bibinfo  {journal}
  {Semicond. Sci. Technol.}\ }\textbf {\bibinfo {volume} {25}},\ \bibinfo
  {pages} {063001} (\bibinfo {year} {2010})}\BibitemShut {NoStop}%
\bibitem [{\citenamefont {Lopes~dos Santos}\ \emph {et~al.}(2012)\citenamefont
  {Lopes~dos Santos}, \citenamefont {Peres},\ and\ \citenamefont
  {Castro~Neto}}]{Lopes:2012}%
  \BibitemOpen
  \bibfield  {author} {\bibinfo {author} {\bibfnamefont {J.~M.~B.}\
  \bibnamefont {Lopes~dos Santos}}, \bibinfo {author} {\bibfnamefont
  {N.~M.~R.}\ \bibnamefont {Peres}}, \ and\ \bibinfo {author} {\bibfnamefont
  {A.~H.}\ \bibnamefont {Castro~Neto}},\ }\href {\doibase
  10.1103/PhysRevB.86.155449} {\bibfield  {journal} {\bibinfo  {journal} {Phys.
  Rev. B}\ }\textbf {\bibinfo {volume} {86}},\ \bibinfo {pages} {155449}
  (\bibinfo {year} {2012})}\BibitemShut {NoStop}%
\bibitem [{\citenamefont {Mele}(2012)}]{Mele:2012}%
  \BibitemOpen
  \bibfield  {author} {\bibinfo {author} {\bibfnamefont {E.~J.}\ \bibnamefont
  {Mele}},\ }\href@noop {} {\bibfield  {journal} {\bibinfo  {journal} {J. Phys.
  D: Appl. Phys.}\ }\textbf {\bibinfo {volume} {48}},\ \bibinfo {pages}
  {154004} (\bibinfo {year} {2012})}\BibitemShut {NoStop}%
\bibitem [{\citenamefont {Mele}(2010)}]{Mele:2010}%
  \BibitemOpen
  \bibfield  {author} {\bibinfo {author} {\bibfnamefont {E.~J.}\ \bibnamefont
  {Mele}},\ }\href {\doibase 10.1103/PhysRevB.81.161405} {\bibfield  {journal}
  {\bibinfo  {journal} {Phys. Rev. B}\ }\textbf {\bibinfo {volume} {81}},\
  \bibinfo {pages} {161405} (\bibinfo {year} {2010})}\BibitemShut {NoStop}%
\bibitem [{\citenamefont {Apalkov}\ and\ \citenamefont
  {Chakraborty}(2011)}]{Apalkov:2011}%
  \BibitemOpen
  \bibfield  {author} {\bibinfo {author} {\bibfnamefont {V.~M.}\ \bibnamefont
  {Apalkov}}\ and\ \bibinfo {author} {\bibfnamefont {T.}~\bibnamefont
  {Chakraborty}},\ }\href {\doibase 10.1103/PhysRevB.84.033408} {\bibfield
  {journal} {\bibinfo  {journal} {Phys. Rev. B}\ }\textbf {\bibinfo {volume}
  {84}},\ \bibinfo {pages} {033408} (\bibinfo {year} {2011})}\BibitemShut
  {NoStop}%
\bibitem [{\citenamefont {Tabert}\ and\ \citenamefont
  {Nicol}(2012)}]{Tabert:2012}%
  \BibitemOpen
  \bibfield  {author} {\bibinfo {author} {\bibfnamefont {C.~J.}\ \bibnamefont
  {Tabert}}\ and\ \bibinfo {author} {\bibfnamefont {E.~J.}\ \bibnamefont
  {Nicol}},\ }\href {\doibase 10.1103/PhysRevB.86.075439} {\bibfield  {journal}
  {\bibinfo  {journal} {Phys. Rev. B}\ }\textbf {\bibinfo {volume} {86}},\
  \bibinfo {pages} {075439} (\bibinfo {year} {2012})}\BibitemShut {NoStop}%
\bibitem [{\citenamefont {Mahan}(1990)}]{Mahan:1990}%
  \BibitemOpen
  \bibfield  {author} {\bibinfo {author} {\bibfnamefont {G.~D.}\ \bibnamefont
  {Mahan}},\ }\href@noop {} {\emph {\bibinfo {title} {Many Particle Physics}}}\
  (\bibinfo  {publisher} {Plenum},\ \bibinfo {address} {New York},\ \bibinfo
  {year} {1990})\BibitemShut {NoStop}%
\bibitem [{\citenamefont {Stille}\ \emph {et~al.}(2012)\citenamefont {Stille},
  \citenamefont {Tabert},\ and\ \citenamefont {Nicol}}]{Stille:2012}%
  \BibitemOpen
  \bibfield  {author} {\bibinfo {author} {\bibfnamefont {L.}~\bibnamefont
  {Stille}}, \bibinfo {author} {\bibfnamefont {C.~J.}\ \bibnamefont {Tabert}},
  \ and\ \bibinfo {author} {\bibfnamefont {E.~J.}\ \bibnamefont {Nicol}},\
  }\href {\doibase 10.1103/PhysRevB.86.195405} {\bibfield  {journal} {\bibinfo
  {journal} {Phys. Rev. B}\ }\textbf {\bibinfo {volume} {86}},\ \bibinfo
  {pages} {195405} (\bibinfo {year} {2012})}\BibitemShut {NoStop}%
\bibitem [{\citenamefont {Ohta}\ \emph {et~al.}(2006)\citenamefont {Ohta},
  \citenamefont {Bostwick}, \citenamefont {Seyller}, \citenamefont {Horn},\
  and\ \citenamefont {Rotenberg}}]{Ohta:2006}%
  \BibitemOpen
  \bibfield  {author} {\bibinfo {author} {\bibfnamefont {T.}~\bibnamefont
  {Ohta}}, \bibinfo {author} {\bibfnamefont {A.}~\bibnamefont {Bostwick}},
  \bibinfo {author} {\bibfnamefont {T.}~\bibnamefont {Seyller}}, \bibinfo
  {author} {\bibfnamefont {K.}~\bibnamefont {Horn}}, \ and\ \bibinfo {author}
  {\bibfnamefont {E.}~\bibnamefont {Rotenberg}},\ }\href@noop {} {\bibfield
  {journal} {\bibinfo  {journal} {Science}\ }\textbf {\bibinfo {volume}
  {313}},\ \bibinfo {pages} {951} (\bibinfo {year} {2006})}\BibitemShut
  {NoStop}%
\end{thebibliography}%

\end{document}